Manuscript title – Word template for Earth Science Reviews ScienceDirect.

**Shannon Entropy Estimator for the Characterization of Seismic-Volcanic Signals using Python.**

Ligdamis Gutiérrez [1, 2] *, Pablo Rey-Devesa [1, 2], Jesús Ibáñez [1, 2], Carmen Benítez[3]

[1] Department of Theoretical Physics and Cosmos. Science Faculty. Avd. Fuentenueva s/n. University of Granada. 18071. Granada. Spain.

[2] Andalusian Institute of Geophysics. Campus de Cartuja. University of Granada. C/Profesor Clavera 12. 18071. Granada. Spain.

[3] Department of Signal Theory, Telematics and Communication. University of Granada. Informatics and Telecommunication School. 18071. Granada. Spain.

Dear Editors-in-Chief,

Please find the enclosed manuscript "**Shannon Entropy Estimator for the Characterization of Seismic-Volcanic Signals using Python.**" By Ligdamis Gutiérrez, Pablo Rey-Devesa, Jesús Ibáñez, Carmen Benítez. Which we are submitting for exclusive consideration for publication in Earth Science Reviews ScienceDirect. We confirm that the submission follows all the requirements and includes all the items of the submission checklist. The manuscript presents:

As far as we know, the authors consider this article groundbreaking. Unlike other similar and more complex tools and systems, it presents a novel approach, especially for learning applications in the efficient and straightforward use of complex metrics with probabilistic mathematical models in the field of volcanic seismicity. Our proposal has four main advantages.

• First, we present a reliable and proven tool that can be deployed in real time to manage large volumes of seismic data from active volcanoes. It performs analysis using four specific probabilistic metrics through a secure and scalable software architecture developed in an open-source language like Python. This provides users with valuable information to improve the analysis and classification of seismic-volcanic signals, which is useful for the future development of efficient early warning models for volcanic eruptions.

- Second, the learning curve and use of software tools by users is quick, in a friendly and visual environment that requires no prior programming knowledge. This allows research to focus on signal analysis, achieving optimal performance in the analysis of seismic-volcanic processes to determine the state of an active volcano in its various stages (pre-eruptive, eruptive, and post-eruptive).

- Third, this work utilizes metrics like Shannon Entropy, Kurtosis, Frequency Index and Energy and methods that have proven to be very effective parameters for quantifying the eruptive state of a volcano. By implementing these metrics from different perspectives, their defining values can be adjusted, allowing for validation during the analysis process of eruptive episodes.

- Fourth, this set of tools meets the basic needs of observatories and seismological research centers that require user-friendly and easy-to-use tools to provide reliable real-time conclusions about the processes and analysis of seismic-volcanic signals. In fact, we believe that the analysis and results of the probabilistic metrics and the computer architecture developed and used in this study, and the associated results, will have a significant scientific and social impact on the communities where they are implemented.

All these advantages have a direct application in the field of analysis and processing of seismic-volcanic signals, which seek, through the use of four advanced probabilistic metrics used in the field of machine learning and AI, to create efficient early warning models for the recognition and prevention of natural disasters caused by volcanic eruptions, which is in line with the spirit of their journal.

The authors confirm that this manuscript nor any part of its content has not been published or presented elsewhere in part or in entirety and is not under consideration by another journal. We have read and understood your journal's policies, and we believe that neither the manuscript nor the study violates any of these. There are no conflicts of interest to declare.

We provide the source codes in a public repository with details listed in the section "Code availability".

Thanks for your consideration.

Sincerely,

Ligdamis Gutiérrez Espinoza. Teaching and Research Staff, (PDI)
University of Granada (Ugr) ligdamis@ugr.es/ ligdamis@yahoo.com

# Highlights

Manuscript title - Word template for Earth Science Reviews ScienceDirect.

**Shannon Entropy Estimator for the Characterization of Seismic-Volcanic Signals using Python.**

Ligdamis Gutiérrez [1,2] *, Pablo Rey-Devesa [1,2], Jesús Ibáñez [1,2], Carmen Benítez[3]

[1] Department of Theoretical Physics and Cosmos. Science Faculty. Avd. Fuentenueva s/n. University of

Granada. 18071. Granada. Spain.

[2] Andalusian Institute of Geophysics. Campus de Cartuja. University of Granada. C/Profesor Clavera 12.

18071. Granada. Spain.

[3] Department of Signal Theory, Telematics and Communication. University of Granada. Informatics and
Telecommunication School. 18071. Granada. Spain.

- Highlight 1
    - Secure, scalable software applications that can be deployed in real-time to handle large volumes of seismic data, enabling the creation of efficient early warning models.

- Highlight 2
    - The standardization of metrics and methods that validate studies during the analysis process of eruptive episodes.

- Highlight 3
    - The learning curve and use of software tools for users must be quick for optimal performance.

- Highlight 4
    - Shannon entropy has proven to be a very effective parameter for quantifying the eruptive state of a volcano.

- Highlight 5
    - In this work, the use of this metric is implemented from different perspectives in which its defining values can be adjusted.

- Highlight 6
    - Observatories and seismological research centers need user-friendly, simple-to-use tools that meet their research needs.

Manuscript title – Word template for Earth Science Reviews ScienceDirect.

**Shannon Entropy Estimator for the Characterization of Seismic-Volcanic Signals using Python.**

Ligdamis Gutiérrez [1,2] *, Pablo Rey-Devesa [1,2], Jesús Ibáñez [1,2], Carmen Benítez[3]

[1] Department of Theoretical Physics and Cosmos. Science Faculty. Avd. Fuentenueva s/n. University of Granada. 18071. Granada. Spain.

[2] Andalusian Institute of Geophysics. Campus de Cartuja. University of Granada. C/Profesor Clavera 12. 18071. Granada. Spain.

[3] Department of Signal Theory, Telematics and Communication. University of Granada. Informatics and Telecommunication School. 18071. Granada. Spain.

Author 1: Ligdamis Gutiérrez, affiliation, ORCID(s): 0000-0002-2585-7006

Author 2: Pablo Rey-Devesa, affiliation, ORCID(s): 0000-0002-6254-4930

Author 3: Jesús Ibáñez, affiliation, ORCID(s): 0000-0002-9846-8781

Author 4: Carmen Benítez, affiliation, ORCID(s): 0000-0002-5407-8335

**Authorship contribution statement.**

Author 1: **Ligdamis, Gutierrez**. Contribution: Conceptualization, Methodology, Software, Validation, Writing - Review & Editing.

Author 2: **Pablo Rey-Devesa**. Contribution: Software, Review.

Author 3: **Jesús Ibáñez**. Contribution: Supervision, Project administration.

Author 4: **Carmen Benítez**. Contribution: Software, Supervision, Project administration.

**ABSTRACT**

Volcanic regions have shaped human settlements for millennia, placing over 500 million people worldwide within close proximity to active volcanoes. Predicting eruptions that threaten both lives and property remains a critical challenge, given the complexity and diversity of seismic signals and recording formats. Effective early warning systems require tools that are not only powerful but also user-friendly, computationally efficient, and capable of handling large seismic datasets and complex mathematical and probabilistic models used in Machine Learning and Artificial Intelligence.

We present a Shannon Entropy Estimator for the Characterization of Seismic-Volcanic Signals using Python designed to meet these demands. Featuring intuitive graphical user interfaces and a low learning curve, the system enables real-time analysis of seismic data using four key metrics: Shannon Entropy, Kurtosis, Frequency Index, and Energy. These metrics provide robust insights into a volcano's state across different stages, independently of the type of event (VT, VLP, Tremor, Hybrid, Explosions) or the data format (SAC, MSEED, SEISAN, etc.). In this work, the efficacy of this tool is demonstrated through the analysis and graphical presentation of these metrics using real seismic records from active volcanoes, providing concrete examples of its practical application.

This approach allows reliable monitoring, accurate categorization, and early detection of critical changes in volcanic activity, supporting the development of predictive models and enhancing the effectiveness of early warning systems. This practical, low-cost solution meets the analytical needs of research institutes and observatories while contributing to the mitigation of volcanic hazards worldwide.

**Introduction**

One of the major needs of most seismic observatories worldwide is to establish and possess reliable technological tools that can facilitate the management of the vast amount of seismic data generated in tectonic and volcanic environments. Many lines of action have been implemented in this regard, through various methods and in different volcanic scenarios. We have seen efforts in diverse and complex volcanoes due to their geological structure, type of eruption, and various other factors. These include different types of seismic events generated by these volcanoes, to which the scientific community must present real-time solutions. These solutions aim to establish increasingly reliable and precise early warning systems.

Over time, the study of monitoring and classifying seismic events in volcanic eruptive processes has proven to be extremely complex, partly due to the structural and geological complexity of volcanoes and the diverse seismic signals they generate during each process. Until a few years ago, the classification and recognition of seismic signals was the main focus of most scientific work aimed at monitoring these processes and establishing efficient and rapid early warning systems for such events. Several authors have proposed in the past the use of techniques such as recognition and classification models based on Hidden Markov Models (Benitez et al., 2006; Gutiérrez et al., 2006; Ibañez et al., 2009; Gutiérrez et al., 2009; Cortés et al., 2021; Bhatti et al., 2016; Trujillo et al., 2018) and neural networks eruptions (Curilem et al., 2009; Scarpetta et al., 2005; Titos et al., 2018; Martinez et al., 2021; Rodríguez et al., 2021), aiming to construct models for forecasting volcanic eruptions.

Currently, to understand what happens in the cycle of an eruptive process (pre-eruptive, eruptive, and post-eruptive), various authors estimate diverse metrics and features to characterize seismic signals in volcanoes during these eruptive periods using artificial intelligence and machine learning techniques (Malfante et al., 2018; Malfante et al., 2017; Lara et al., 2020; Ren et al., 2020; Falcin et al., 2021).

Recent studies propose the use of seismic features like Shannon Entropy, Kurtosis, Frequency Index and Energy among others, as a surveillance element with forecasting projection Rey-Devesa et al. (2023 a,b). Entropy features (Konstantinou et al., 2022; Telesca et al., 2014; Boué et al., 2015), Kurtosis (Baillard et Al., 2014; Li, X et al., 2016; Ma, H. Quian et al., 2019; Langet et al., 2014) and features related to the energy content recorded in different frequency bands (Buurman et al., 2006; Ketner et al., 2013; Ardid et al., 2022) have been studied in works with seismic records in volcanoes before and after eruptions, determining better monitoring of these events to establish reliable patterns for determining early warning models in eruptive activity.

In this work we propose the use of an innovative software designed for the exhaustive analysis of seismic record, capable of reading different recording formats and filtering the signal for removing outliers and avoiding the noise associated to the desired frequency bands. The analysis implements a moving window of a certain size that can be adjusted depending on the data density, the length of the record or the targeted event. Moreover, spectral decomposition capabilities enable the examination of seismic signatures across various frequency bands, facilitating the discrimination of source processes. This method allows the seismic variability during unrest periods of activity (eruptions) and inactivity (pre- and post-eruptions) by analyzing the trend of its characteristics through the combination of the four metrics proposed here (Shannon Entropy, Kurtosis, Frequency Index, and Energy), providing valuable information on volcanic dynamics. The graphical representation and optimized data presentation mechanisms facilitate the understanding of the results. Together, these features constitute a robust and efficient tool with great potential for predicting volcanic eruptions, thus contributing to risk management and the protection of human life and property.

The implementation of these sets of tools in real-time is a simple task considering the nature of the extracted features that do not require a high computational power. The ability of this software to perform exhaustive and systematic analysis using adjustable moving windows and spectral decomposition is based

Ligdamis Gutierrez: Preprint submitted to arXiv. 7

on efficient algorithms. This allows that the calculations can be easily and rapidly operated even at environments with limited computational resources. Thus, we propose this tool for its integration in real-time monitoring systems, which can estimate crucial information for volcanic eruption forecasting and risk management.

**1.1. - Methodology**

Of the three modules that make up this set of analysis tools, the first focuses on identifying which component is most suitable for use in the analysis. This module benefits from signal filtering, allowing for a better assessment of which component and at which station the identification of seismic events is most effective for analysis.

The third module involves a new analysis, this time focusing on selecting the type of filter to use in order to establish the envelope that best determines the calculation in Shannon Entropy.

It is in the second module where the process of calculating and analyzing the mathematical metrics to be used with seismic signals takes place.

The set of mathematical tools used in this article is based on four main parameters or metrics. These parameters include precise mathematical and probabilistic calculations that provide significant value in assessing the complexity and structure of seismic signals present in active volcanoes.

Specifically, these are:

a) Shannon Entropy,
b) Kurtosis,
c) Frequency Index and
d) Energy.

This section briefly describes their importance in the use of seismic signals during the stages of eruptive processes (pre-eruptive, eruptive and post-eruptive) and explains why they were chosen for the calculations in the seismic analysis process that is described in Module Two of this study.

### a) Shannon Entropy

Shannon Entropy H(*x*), is a concept proposed in 1948 by mathematician Claude Shannon (Shannon, C. E. 1948). It states that: *"Entropy measures the amount of uncertainty or randomness in a dataset or signal,"* in this case, seismic signals.

Seismic signals from activity in active volcanoes contain information about various types of geophysical events generated by volcanoes, such as earthquakes, explosions, tremors, and long-period events, among others.

By analyzing seismic signals through Entropy calculations, an increase in Entropy values indicates greater uncertainty or disorder in the data. This suggests that the presented seismic information becomes significantly more difficult to predict.

Therefore, when Entropy is at its maximum, it indicates greater randomness. Conversely, lower Entropy values suggest reduced uncertainty.

Shannon Entropy H(*x*) is defined by the following formula (1):

$$H(x) = -\sum_{i=1}^{n} p(x_i) \log_2 p(x_i) \qquad (1)$$

Where Shannon Entropy H(x), for a random variable (*x*), takes different values from 1 to (*n*), with probabilities *p(x_i)*. In this way, Entropy is related to the probabilities of occurrence of the various possible values of (*x*).

If all probabilities are equal, the Entropy value reaches its maximum, indicating greater randomness. Conversely, if there is a dominant value with a probability higher than the others, the resulting Entropy value will be lower, reflecting reduced uncertainty.

In the case of seismic signals, the Entropy value provides a measure of the complexity and randomness in the time series of the various seismic waves generated by active volcanoes.

Ligdamis Gutierrez: Preprint submitted to arXiv. 9

On the other hand, Entropy can also be used to identify sudden changes in the properties of seismic signals. This is particularly useful for detecting transitions in seismic signal behavior before, during, and after seismic events such as swarms or eruptions.

For example, higher-frequency events like earthquakes or explosions often show a significant reduction in Entropy values. This occurs due to the inherently more organized or coherent nature of the seismic waves generated by such events, compared to background noise from environmental factors or external sources like human or animal activity in the area.

This leads to another highly useful aspect of evaluating Entropy results: analyzing seismic signals to distinguish between earthquake-type events, microseisms, and anthropogenic events, such as those caused by human activity or local wildlife.

By doing so, it becomes possible to differentiate between earthquakes originating from the volcano under study and microseisms from external sources. This distinction is evident in the Entropy values, as seismic signals from earthquakes typically exhibit much lower Entropy within certain frequency ranges, whereas events of external origin or background noise tend to have higher Entropy due to their greater randomness.

Therefore, by analyzing Entropy values in seismic signals over extended periods—months or even years—it is possible to detect both the onset and conclusion of seismic events with much greater precision.

This highlights how Entropy can be used to analyze so-called *"premonitory"* signals or events. The Entropy value in seismic signals may drop to near-zero levels before major events, such as a volcanic eruption. This phenomenon can aid in predicting both strong earthquakes and imminent eruptions in active volcanoes by identifying changes in the dynamics of geophysical forces within seismic waves.

In this sense, for example, metrics and models based on Shannon Entropy, such as the Fisher-Shannon method (Telesca et al. 2014) and Permutation Entropy (PE) (Konstantinou et al. 2022), have used calculations with statistical tools such as Shannon Entropy, in seismic signals prior to and during eruptive episodes, as well as seismic noise monitoring in order to detect changes in the dynamics of a volcano and to be an aid in preventing new eruptive episodes.

As a result, the Entropy parameter provides a highly rigorous mathematical approach both for analyzing large volumes of seismic records and as a useful tool in studying changes in the dynamics of a volcano. It can also be a factor in the prevention of eruptive episodes, thanks to its ability to quantify the uncertainty in seismic signals.

**b) Kurtosis**

Kurtosis (**Cr**) is a statistical parameter (DeCarlo, L. T. 1997) that describes the shape of the distribution of information contained in records or datasets, particularly regarding the *"sharpness or energy peaks"* or *"flatness"* of the distribution compared to a normal (Gaussian) distribution.

Mathematically, Kurtosis is defined as the fourth central moment of a distribution divided by the square of the variance. It is related to the degree of concentration of energy in central or extreme values.

In seismology, Kurtosis measures the shape of a seismic signal's distribution, particularly whether it has sharp peaks (i.e., events with high concentrated energy) or a flatter, more uniform distribution.

Unlike Entropy, which quantifies randomness, Kurtosis focuses on extreme values and the shape of the signal around its mean. This makes it particularly useful for locating seismic events. Specifically, it can be applied to the following aspects:

*i) Anomaly Detection:*

Kurtosis can be used to detect anomalies or unusually occurring events in seismic records, particularly in active volcanoes. A sudden increase in the calculated Kurtosis value may indicate the occurrence of a significant seismic event, helping to distinguish it from background noise.

*ii) Identification of Transient Events:*

Seismic signals—such as earthquakes, explosions, or long-period (LP) events—are often characterized by abrupt changes over time. When a seismic event occurs, its energy is concentrated over short time periods, resulting in a distribution with **high Kurtosis** (*leptokurtic*), meaning its value is greater than three. In this case, the distribution exhibits a sharper peak and longer, thicker tails, indicating a higher frequency of extreme values (both peaks and coda waves).

Conversely, background signals tend to be more uniform and continuous, leading to **low Kurtosis** (*platykurtic*), typically with a value lower than three. This results in a flatter central distribution and less pronounced coda waves, meaning extreme values are less frequent.

When Kurtosis values are close to three, the distribution is referred to as mesokurtic, meaning it behaves similarly to a normal distribution, without extreme peaks or tails.

*iii) Characterization of Signal Shape:*

Kurtosis can also help assess the shape of a seismic signal and its propagation. Seismic signals with very sharp peaks and pronounced coda waves typically have **high Kurtosis** values.

This can result from different seismic source mechanisms or propagation effects, such as the nonlinearity of the medium through which seismic waves travel from their source. Understanding these variations helps in interpreting the nature and behavior of seismic activity more accurately.

Thus, when an earthquake occurs in an active volcano, the energy of the generated seismic wave is distributed unevenly, concentrating in specific peaks. This results in an increase in the Kurtosis value of the seismic signal.

Analyzing this parameter allows for distinguishing and identifying true seismic waves originating from a specific seismic event within the Volcano from other waves generated by environmental noise sources or external anthropogenic vibrations.

The **Kurtosis (*Cr*)** is calculated using the following formula (2):

$$Cr = \frac{\sum_{i=1}^{n}(-\bar{x})^4}{n \cdot S^4} - 3 \qquad (2)$$

Where:

- **[*i*]**: is the amplitude of the seismic signal at sample "*i*".

- $\bar{x}$ : is the mean of the signal.
- *S*: is the standard deviation of the signal.
- *n* : is the total number of samples.

The calculation process determines the following:

- The **mean** of the seismic signal is calculated (i.e., the average of all signal values).

- The **summation** $\sum$ accounts for the deviation of each data point $x_i$ from the mean ($\bar{x}$).

- The fourth central moment (numerator) evaluates how the seismic signal values are distributed relative to the mean, quantifying the dispersion of values raised to the fourth power.

- The denominator consists of the variance squared, normalizing the measure.

In Formula **(2)**, an adjusted Kurtosis is used by subtracting (-**3**) from the result, so that a normal distribution has a Kurtosis of zero. This adjustment makes interpretation easier:

- ***Kurtosis ≈ 0*** → the distribution is similar to a normal (Gaussian) distribution (*mesokurtic*).

- ***Kurtosis > 0*** → the distribution has more pronounced peaks and heavier seismic coda waves. (*leptokurtic*).

- ***Kurtosis < 0*** → the distribution is flatter with lighter seismic coda waves (*platykurtic*).

This formula helps quantify whether a seismic signal has extreme peaks (high Kurtosis) or is more evenly distributed (low Kurtosis).

On the other hand, the difference between Shannon Entropy and **Kurtosis** lies in their focus:

- **Shannon Entropy** measures the complexity**,** uncertainty**, or** randomness of a given seismic signal or seismic record.

- **Kurtosis**, in contrast, focuses on extreme values and the shape of the signal around its mean, making it useful for identifying high-energy, short-duration seismic events.

**c) Frequency Index**

The third parameter analyzed in this study is the **Frequency Index**.

In seismology, the **Frequency Index** is a valuable metric for analyzing the spectral content of seismic signals. It represents the relationship between the different frequencies that compose a given seismic signal and their evolution over time.

This parameter is particularly useful for identifying and classifying different types of seismic events (Wassermann, J. 2012) related to volcanic activity, especially during eruptive processes.

In general, the **Frequency Index (FI)** is a calculation process that quantifies whether a given seismic signal predominantly contains **high** or **low** frequencies.

Its definition is based on the difference or relationship between the dominant frequencies of a seismic event within two spectral ranges.

The **FI** was first calculated for calibration events and then applied to all earthquakes during the swarm, following the procedure described by Buurman & Wets **(**Buurman and West 2010**)**.

The **Frequency Index** is calculated using the following formula (3):

$$FI = log \frac{(A_{high})}{(A_{low})} \qquad (3)$$

Where **A** is the mean amplitude of the seismic waveform within a high or low-frequency band.

The FI is useful for monitoring eruptive processes because:

- **Low FI values** (frequencies below 1–5 Hz) suggest the movement of magma within the volcano, the accumulation of pressurized gases, or the triggering of initial explosions with large volumes of magma in motion.
- **High FI values** (frequencies above 5 Hz) indicate rock fractures caused by magma pressure or explosive volcanic activity, where abrupt energy release generates high-frequency seismic signals.

This formula helps determine whether the energy distribution in a seismic signal is dominated by high or low frequencies, which is key for classifying volcanic seismic events.

The usefulness of the **Frequency Index (FI)** in seismic signals during eruptive processes lies in its ability to distinguish between different types of volcanic activity.

If seismic signals are dominated by **low frequencies** (typically below **1–5 Hz**), they are often associated with:

- **Volcanic tremor** – a continuous low-frequency vibration caused by prolonged movements of magma, gases, or fluids within the volcanic edifice.
- **Long-period (LP) events** – low-frequency seismic signals related to the movement of fluids (magma or gases) in cavities or conduits inside the volcano.

A **low Frequency Index (FI)** may indicate:

- **Movement of magma** at deep levels within the volcanic edifice or its ascent toward the surface.
- **Accumulation of pressurized gases**, which are close to being released violently, possibly signaling an **imminent eruption**.
- **Triggering of initial explosions**, which generate events dominated by low-frequency energy due to the large volume of moving magma or material.

On the other hand, **high Frequency Index (FI) values** indicate the presence of high-frequency signals (above **5 Hz),** which may suggest:

- **Rock fracturing** in response to the pressure exerted by ascending magma.

Ligdamis Gutierrez: Preprint submitted to arXiv. 15

- **Volcanic explosions**, which are abrupt events generating high-frequency seismic signals due to the sudden release of energy within the volcano.

During a volcanic eruptive process, seismic signals typically begin with low frequencies, indicating the ascent of magma from the magma chamber toward the surface.

As magma approaches the surface, internal pressures within the volcano increase, leading to rock fracturing or explosive events. This results in a greater predominance of high frequencies in seismic signals, causing the Frequency Index (FI) to rise significantly**.**

Therefore, the **Frequency Index (FI)** is a valuable analytical tool that helps determine the predominance of high or low frequencies in seismic signals, allowing for a better understanding of volcanic activity.

By providing insight into internal volcanic processes; such as magma movement **(low frequency)** or rock fracturing and explosions **(high frequency)**, this parameter enhances the monitoring of seismic events associated with eruptions. As a result, it contributes to the development of more effective early warning systems for imminent volcanic eruptions, improving risk management and preparedness.

**d) Energy**

The fourth and final parameter included in this work is Energy. Energy is very important in signal processing. In seismology, it allows for the characterization of earthquakes and volcanic explosions. It is also useful for distinguishing between different types of seismic events and for effectively monitoring the evolution of volcanic activity. This makes it possible to detect and anticipate changes, and thus determine more accurately the state of the volcano's evolution over time, evaluate the risk, and make informed alert decisions.

The energy of a seismic signal represents the amount of work done by seismic waves as they propagate through the Earth. The energy of a seismic signal "x(t)" in the time domain is obtained by integrating the square of the signal's amplitude over time. The Energy is calculated using the following formula (4)

$$E = \int_{-\infty}^{\infty} |x(t)^2| dt \qquad (4)$$

This integral represents the area under the curve of the squared magnitude of the signal in the time domain. However, in practice, with digital signals (*such as those recorded by seismometers*), the discrete version is used, and its formula is as follows. (5):

$$E = \sum_{i=1}^{N} x_i^2 \qquad (5)$$

Where: $x_i$ It is the amplitude of the seismic signal at a given point in time (*i*).

**N** It is the total number of samples of the signal.

In this case, the energy of the signals is calculated by summing the squares of the amplitudes, which allows seismic events to be distinguished from noise. This makes it possible to identify the peaks of the events present in the signal (VT, Explosions, LP).

The usefulness of calculating this parameter lies in the fact that it allows for the calculation of the total energy contained in a seismic signal within a specific time interval. In this way, it becomes possible to evaluate the frequency ranges that carry the most energy, allowing for the differentiation between low- and high-frequency signals. This, in turn, can provide information about the origin of the seismic events— whether they are caused by magmatic activity or by rock fracturing due to pressure or gases.

In this work, the discrete formula mentioned in (5) will be used to calculate the energy present in the seismic signals.

Now, in the analysis of the energy calculation in seismic signals, its resulting value can be interpreted as follows:

- High Energy Values: These may indicate seismic events with high energy content, such as earthquakes or volcanic explosions.

- Low Energy Values: These could be associated with exogenous or environmental noise signals. They could also indicate minor tectonic activity that does not represent an imminent eruption.

- Energy Variations: The presence of variations over time could indicate changes in seismic activity, such as magma ascent prior to an eruption or a collapse of the volcanic structure following an eruptive event.

Therefore, the analysis of energy, combined with parameters such as Entropy, Kurtosis, and the Frequency Index, provides an extremely useful tool, both for monitoring and, most importantly, for the evolution and prediction of seismic and volcanic activity. This helps in preventing imminent eruptions and enables the development of better predictive models for volcanic eruptions. This leads to the following conclusions.

**e) Final conclusions.**

In this study, the combined calculation of **four key metrics**—Shannon Entropy, Kurtosis, Frequency and Energy Index—proves to be highly useful for monitoring **seismic activity** throughout the **pre-eruptive, eruptive, and post-eruptive** stages of a volcano.

Each metric provides valuable insights at different stages:

i. **Pre-eruptive Phase**

- **Shannon Entropy H(x)**:

Increases as the complexity of the volcanic system rises. The ascent of magma generates more variable and unpredictable seismic signals, leading to higher entropy values.

- **Kurtosis (Cr)**:

Observed values may be low to moderate, as abrupt seismic events (such as explosions) are not yet occurring. However, deep rock fractures within the volcanic edifice may create occasional peaks.

- **Frequency Index (FI)**:

Low-frequency dominance is observed, indicating deep magma movement and the accumulation of pressurized gases, which could signal the early stages of volcanic unrest.

- **Energy (E)**:

Energy may show a progressive increase, which would indicate the accumulation of pressure due to rising magma, generating fractures and tensions. This leads to seismic events that tend to release more energy as pressure increases (VT, Explosions, and Earthquakes). This increase in seismic energy could also indicate that the volcanic conduit is opening up, signaling rock fracturing. Additionally, changes in released energy, especially if a sudden or sustained increase occurs, would indicate the imminence of an eruptive episode.

ii. **Eruptive Phase**

- *Shannon Entropy H(x)*:

During an eruptive phase, the value of this metric tends to decrease drastically, approaching zero, as seismic signals become more repetitive and predictable due to successive explosions. Thus, a drastic decrease in Entropy indicates an eruption. At this point, it is important to determine, through successive analyses, when this decrease begins. This would allow us to predict how far in advance eruption alerts should be issued to safeguard the lives of those in the surrounding area.

- *Kurtosis (Cr):*

High values are expected at this stage, as the explosions generate sharp energy spikes in the seismic signals. This is a key indicator of the occurrence of violent eruptive processes at the volcano.

Ligdamis Gutierrez: Preprint submitted to arXiv. 19

- *Frequency Index (FI):*

High values are expected during the eruptive phase because seismic signals during eruptions are determined by high frequencies, resulting from rock fracturing and surface explosions. A rapid increase in high-frequency peaks in this metric indicates a shift toward a highly explosive eruptive phase.

- *Energy (E):*

An increase in this indicator could signal various processes: it could reflect explosive pulses, magma movement, or crater collapse. Furthermore, it is important to consider that a volcano can experience different eruptive processes (Caldera, J., & Wirasinghe, S. C. 2014), both effusive and explosive, which generate varying levels of seismic energy. Violent eruptive episodes tend to release more energy in a short period. This will determine the predominant type of eruption at the volcano.

iii. **Post-Eruptive Phase**

- *Shannon Entropy H(x):*

In post-eruptive stages the value of this metric generally increases again, which happens because the system gradually becomes more chaotic due, for example, to internal factors in the volcanic edifice such as: the collapse of volcanic conduits and random emissions of fluids in the volcano (e.g., gas releases).

- *Kurtosis (Cr):*

A decrease in the value of this metric indicates fewer energy peaks and more homogeneous seismic activity.

- *Frequency Index (FI):*

If low frequencies predominate, it indicates that the magma has stopped rising and that the volcanic system is cooling down.

- *Energy (E)*:

In the post-eruptive stage, the energy value following an eruption is typically expected to gradually decrease. If this does not occur, it could indicate an imminent eruptive reactivation or residual activity. Furthermore, elevated values could be related to the collapse of the crater or volcanic edifice, or possibly to a readjustment or reorganization of its internal structure.

As has been demonstrated, each of these parameters provides a different yet complementary perspective on the nature of seismic signals. When integrated, as presented in the study modules, they not only allow for more precise monitoring but, more importantly, enable the anticipation of changes in volcanic activity. This, in turn, facilitates more accurate and reliable predictions regarding the future evolution of eruptive processes.

This approach adds significant value to the experience of seismic observers by providing mathematical tools that yield reliable results, enabling more precise analysis and decision-making in response to potential volcanic events. Ultimately, it contributes to the development of more effective early warning systems for volcanoes, helping to save lives and mitigate the associated risks.

Finally, within the branch of Physics that studies the Earth and its physical phenomena, which is Geophysics, we believe that the methodology of our work "Shannon Entropy Estimator for the Characterization of Seismic-Volcanic Signals using Python" can be included as a study of "Computational Geophysics".

1. **Software Designation.**

Many seismological and volcanological research observatories use software packages or proprietary software for volcano-seismic events analysis and calculations. An example of this is Earthworm (http://folkworm.ceri.memphis.edu/ew-doc/), which is based on Java. Another example is GeoVista Studio. Takatsuka, M., & Gahegan, M. (2002), which is based on LabVIEW (National Instruments Corp., 2001). These packaged software solutions are quite complex, making it challenging for many users and researchers without programming experience to operate, especially when modifications to the code are needed.

On the other hand, the vast majority of developers have used graphical environment languages like Matlab® (The MathWorks Inc., Massachusetts, USA) for generating codes and programming routines that perform all the calculations and data analyses mentioned earlier. One advantage of Matlab® is that it is an independent language with a relatively simple syntax and contains a large number of built-in high-level calculation and analysis libraries. However, one of its disadvantages is that it is not free or open-source language; it is paid, and licenses are expensive for individual users or research entities; it does not have open-source code, meaning that those who use it, in this case, the scientific community, have to rely on the customer service provided by the company. Examples of the results of using Matlab® in the analysis of seismic-volcanic signals are presented in several works (Curilem et al., 2014; Lesage Philippe, 2009; Messina, A., & Langer, H., 2011; Minio et al., 2023).

Another language that has become one of the most widely used by developers for routines and code analysis and calculations for seismic events is Python. One advantage of Python over Matlab® is that it is cross-platform; it can be used on various operating system environments, such as Windows, Linux, and Mac. Python syntax is similar to Matlab's, and being open-source allows for easier debugging of errors and necessary changes by the user. Additionally, Python has a large volume of libraries that are constantly growing thanks to a vast community of developers, facilitating the use of many complex techniques and routines. Among the works used for seismic calculations and analysis are those by Bueno et al. (2020), Lecocq et al. (2014) and Smith et al. (2020). Modules in Python have even been developed for automatic detection of seismic events in volcanoes. Fenner et al. (2021), for event localization based on seismic amplitudes. Cornejo-Surez et al. (2018), or for analyzing volcanic eruption predictions using characteristic seismic data present in volcanoes. Killion et al. (2018). Another noteworthy Python application for seismic data analysis and processing is the library Obspy, which offers a suite of tools for advance seismic signal analysis, like accessing to seismic databases, preprocessing raw data, visualizing waveforms, detecting seismic events and conducting spectral analysis among others. Beyreuther et al. (2010).

However, both the early works and the more complex ones in spectral analysis, event detection, and classification present a development and software management environment, as well as programming languages that are quite rigorous in terms of design and complexity. In some cases, this entails a learning curve, adaptation, and equally complex updating. Many of these software packages have been determined and developed to address a specific problem or analysis within the broad realm of studying seismic signals in volcanoes. This leaves the use of the tools incomplete and adds the problem of having to learn the language and develop new routines to complement the research or studies required in observatories or during a specific volcanic eruptive process. Another aspect of the complexity arises from the difficulty of transitioning from an older version of the language to a much more recent one in which the codes are developed. This entails difficulties for users who are generally not trained to make complex modifications to the software packages and codes they handle.

Therefore, in this work, we offer a set of computer tools that are quite fundamental, simple and very fast to compute. Yet, they have been proved to be essential and important in the applications of signal processing to the analysis of seismic. These tools consist of a single module, which is quite simple to learn and use, as well as reliable and robust. They are written in Python language and have a very user-friendly graphical user interface (GUI). The user only needs to input general data, without worrying about the calculations and analyses that the system will perform. They only need to press specific command buttons on the interface for the task or function required, and thus obtain the desired results. Modules that are highly scalable and can be easily adapted, modified, and updated to larger and more comprehensive versions individually without affecting the rest of the modules. This set of tools developed in Python for calculating metrics such as Shannon Entropy, Kurtosis, Frequency Index and Energy, is useful for complex analysis of a large number of seismic-volcanic signals (up to one year per seismic record calculation). The calculation and definition of these features is described in Rey-Devesa et al. (2023 b). It represents a fairly accessible, simple, and useful option for most observatories and their users to reliably determine pre- and post-eruption scenarios in volcanoes and to establish risk scenarios and early warning models for future eruptions with greater accuracy. In this way, works like the one mentioned above confirm the standardization of metrics and methods that validate studies during the analysis process of eruptive episodes. This serves as a solution to the lack or absence of simple and practical tools for many of the various analyses carried out in observatories with seismic-volcanic signals.

This set of tools has been tested in Windows, Linux, and Mac environments, as well as with various datasets from different volcanoes. Therefore, their reliability, ease of use, and optimal results have been verified.

### 1.1. Toolkit Description

This set of software applications consists of three individual tools or modules that, in a user-friendly manner, provide calculation and analysis methods through graphical user interfaces (GUIs). In addition to these three modules, the system begins with a welcome or home module, which includes access to each of the working modules (Cfr. Fig 1). Furthermore, the system includes a user manual written in Spanish and English, as well as basic instructions for installing the libraries and Python software needed for the system to function properly.

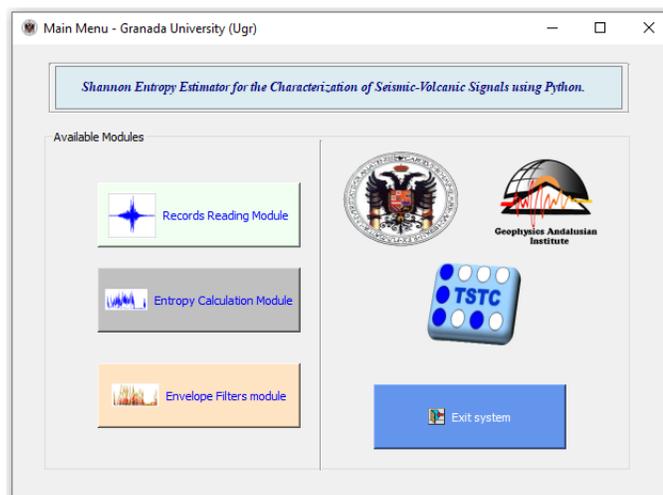

Figure 1: Module 1: welcome or home module. It provides access to the three working modules of the seismic-volcanic signal characterization system, indicated by the command buttons.

The three analysis tools (modules) described in this work are the following:

1. - Performs the read, filtering and plot the seismic signals.

2. - Entropy estimator for Characterization of Volcanic Seismic Signals.

3. - Performs the read, filtering and plot the entropy envelope with various frequencies.

Among the main functionalities and utilities that this set of tools provides to researchers and observatories, the following can be mentioned:

(a) Reading seismic records in various formats commonly used in national and international seismological centers (MSEED, SAC, WAV, GSE2, EVT, GCF, among others).

(b) Performing commonly used analyses in observatories, involving the use of various digital filters with one or three components (North-South, East-West, and Vertical).

(c) Calculating Shannon Entropy, Kurtosis, Frequency Index and Energy with their respective smoothed function, using analysis windows of 5 and 10 minutes, one hour, or 24 hours. Applying different smoothing filters with width sizes such as 50, 100, 300, etc.

(d) Comparing, calculating, and plotting Shannon Entropy smoothed function using various frequency values of a specific filter, defined by the user.

At the same time, the possibility to present the results of the analyses graphically and store them in various formats (PNG, JPG, GIF, and SVG, among others) is offered. Each of the modules includes various data input validation windows to check for user input errors and prevent the system from crashing as a result.

## 2. Toolkit Overview

### 2.1. Performs the read, filtering and plot the seismic signals. (Module 1)

This first work module comprises a user-friendly interface that, through incorporated libraries, supports the reading of various seismic formats used in observatories and seismological institutes such as SAC, MSEED, GSE2, EVT, GCF, WAV, among others.

It allows for easy and efficient management of reading and plotting seismic record graphs using one or three components. Prior to spectral analysis, various digital filtering techniques can be applied to seismic signals, such as Lowpass, Highpass, Bandpass, and Bandstop.

Additionally, the module provides the ability to store graphical results in various formats, such as PNG, JPG, EPS, PS, PDF, RAF, TIF, among others.

The module includes spectral and filtering analysis of records with multiple traces, such as records stored in "MSEED" format, for example. The user can determine which trace to analyze and present results from a record containing more than one trace. General statistics information of the record, as well as metadata information and the number of traces composing the record, are included.

The main interface of the module, presenting the data inputs, types of filters, and spectral analysis, is shown in Figure 2. The result of the calculation process of this module with three components is shown in Figure 3.

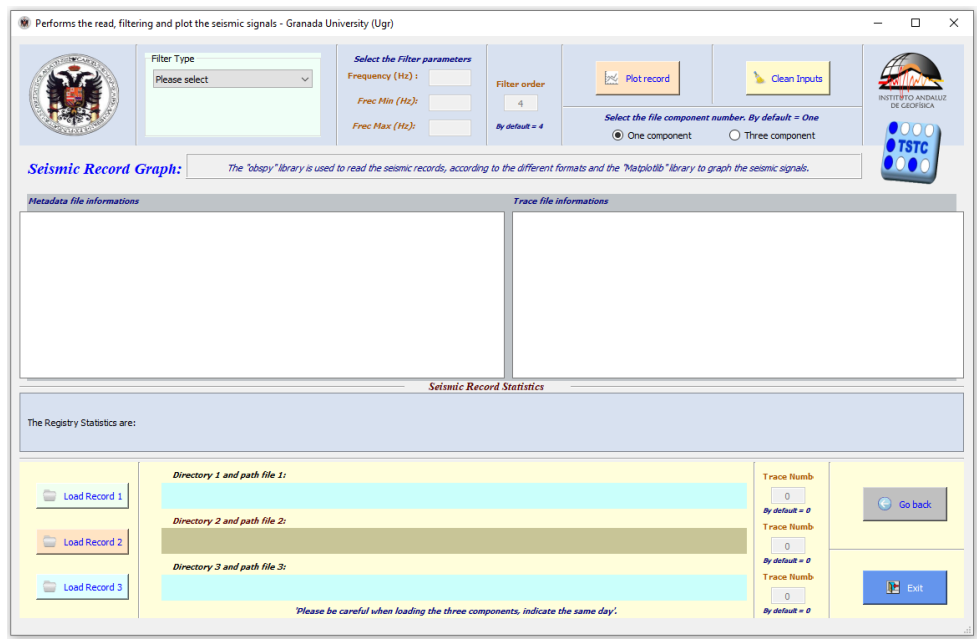

Figure 2: Module 1: Seismic Record Reading and Plotting with One or Three Components.

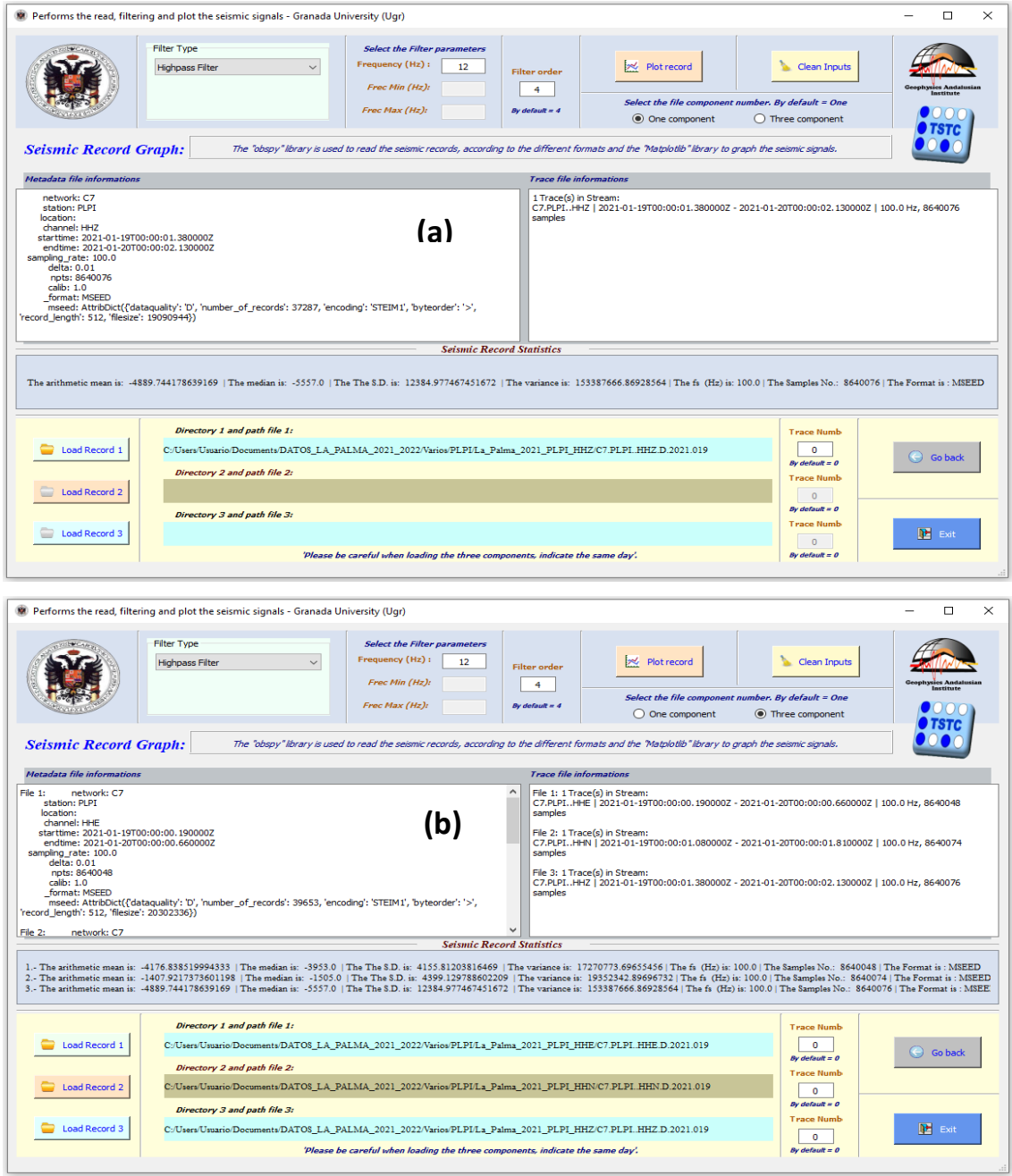

Figure 3: Results of the Module 1 for reading and plotting seismic records using: (a) one component (**b**) all three components.

In the figure 3, the initial default parameter values are determined with a filter order = 4, with the analysis selection of one component. This indicates that when the filter type is selected, only the first record is activated, setting the initial trace to zero, which is the default setting. If the user selects to analyze all three components, the load buttons to select three records with their corresponding traces are activated. Once the filter parameter values to be chosen are entered, simply clicking the 'Plot record' button will display the resulting graph.

The 'Clean Inputs' button clears the entries, first presenting a warning window that prompts the user to confirm whether they really want to clear the entries. If confirmed, the entries will be deleted, and the interface will return to the initial values. If not, they will remain in the interface for further analysis. The process of representing the three seismic components (North-South, East-West, and Vertical) of a seismic record is useful for determining which component to use for mathematical calculations and analysis with the next two modules. This process highlights the true simplicity and ease of use of this and the next two interfaces for the user.

The, Figure 4 shows the user selecting one and three-component record. The command buttons to load these records from the selected folder are activated. Each record has a single trace, which is by default trace = 0. A band-pass filter between 1 and 15 Hz is chosen, and the filter order is left at the default value of 4. After doing this, the user clicks the 'Plot record' command button to see the results. The graph of this analysis is shown in the results section 4.1.

NOTE: The supplementary documentation, as well as the software comprising this module, can be accessed and downloaded from the link provided in "Code availability section".

### 3.2 Entropy estimator for Characterization of Volcanic Seismic Signals. (Module 2)

This second module constitutes a user-friendly interface that allows for easy and efficient management of the calculation process and representation of the metrics of Shannon Entropy, Kurtosis, Frequency Index and Energy. The initial graphical interface of the module is presented in Figure 4.

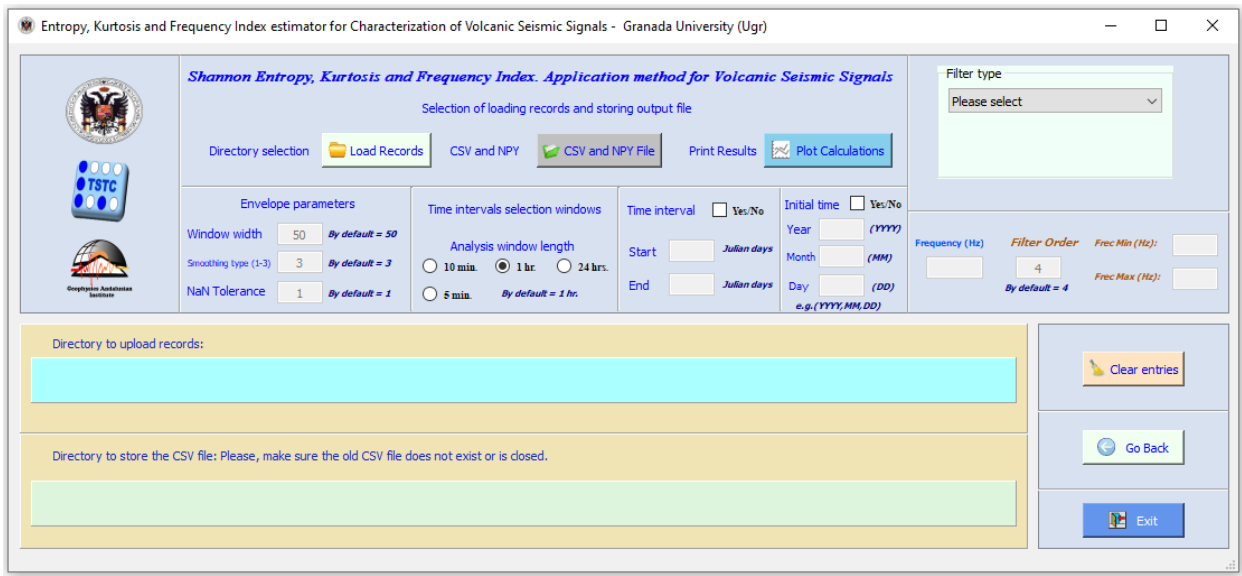

Figure 4: Module 2: Seismic Record Reading and Plotting using Shannon Entropy, Kurtosis, Frequency Index and Energy metrics.

This module reads the folder containing the seismic records to be analyzed. These seismic records can either span the entire year or, by using the "Time interval" block, specify the start and end times for analysis, expressed in Julian days. A general feature of the modules comprising this system is the ease of reading various seismic formats, which, through incorporated libraries, access diverse seismic formats commonly used in observatories and seismological institutes such as SAC, MSEED, GSE2, EVT, GCF, WAV, among others. It presents two directory blocks, one where the path to the folder containing the records will be shown, and the other where two result files in CSV and PY formats will be stored.

The "Envelope parameters" block indicates the parameters of the values that the smoothing filter will take when calculated (size of the filter, type of smoothing, and tolerance for NaN or null values), which are initially set to default values of (50, 3, and 1) respectively.

The "Time intervals selection Windows" block indicates the size of the analysis window, which is initially set to one hour but can be selected as 5 and 10 minutes or 24 hours.

The "Initial time" block indicates the start time of the data in the Gregorian calendar (year, month, and day) for graphical representation on the "x" axis in the output results.

The types of filters for analysis are, as indicated in the previous section: Lowpass, Highpass, Bandpass, and Bandstop. When selecting a particular type of filter, the corresponding input value boxes are activated. The default value for the filter order is four. Another feature similar to the previous module is the ability to store graphical results in various formats such as PNG, JPG, EPS, PS, PDF, RAF, TIF, among others. An additional aspect in the graphical results is the ability to zoom in on each of them to better visualize the records over time.

Once the values are entered, the result observed in the interface is presented in Figure 5. The graphical results of this analysis are shown in Results section 4.2.

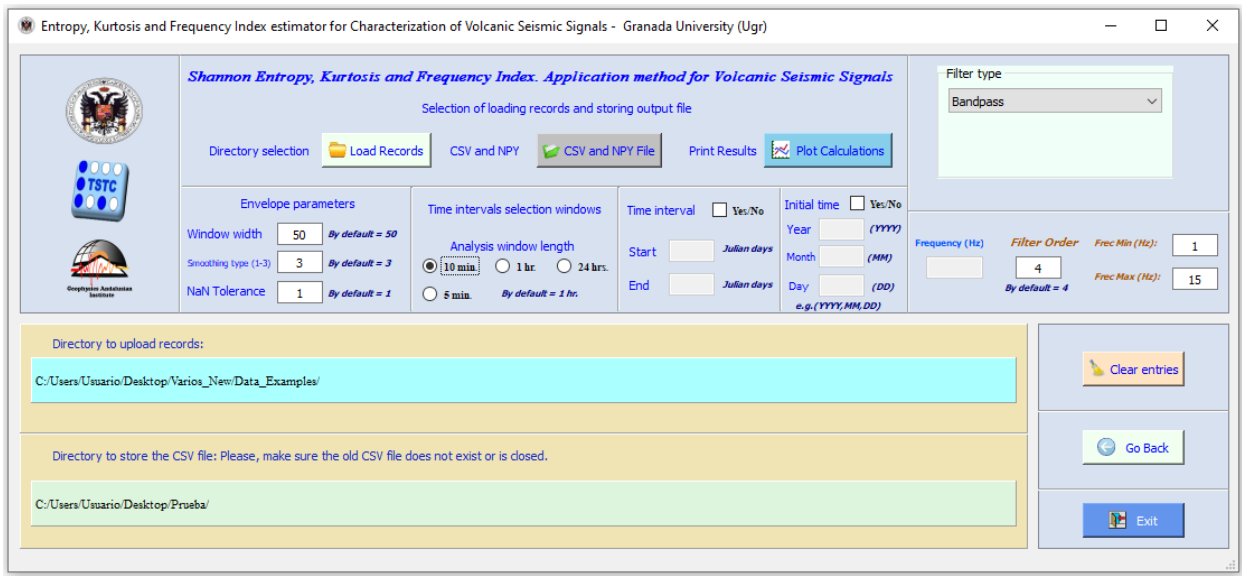

Figure 5: Results of the Module for reading and plotting seismic records using Shannon Entropy, Kurtosis, Frequency Index, and Energy metrics.

NOTE: The supplementary documentation, as well as the software comprising this module, can be accessed and downloaded from the link provided in "Code availability section".

## 3.3 Performs the read, filtering and plot the entropy envelope with various frequencies. (Module 3)

This third module constitutes a user-friendly interface that allows for easy and efficient management of the calculation, comparison, and graphical representation of the Shannon Entropy envelope, calculated at different values according to a specific filter. The initial graphical interface of the module is presented in Figure 6.

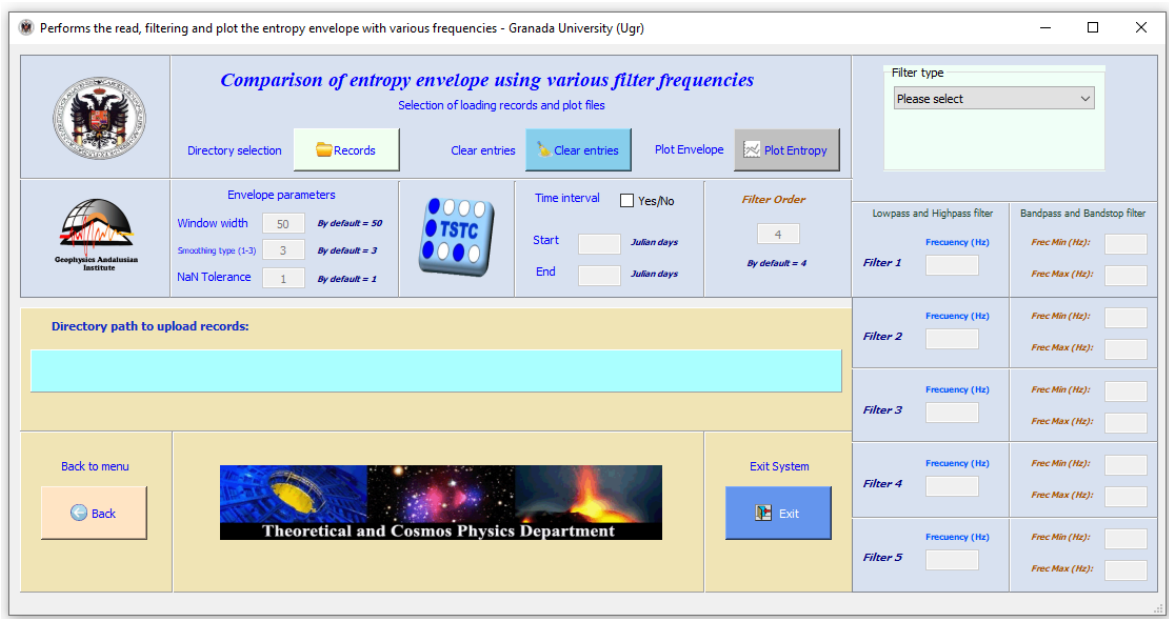

Figure 6: Module 3: Reading, filtering, and comparison of five values of a specific seismic record filter.

Similarly, to the previous modules, this module loads the folder containing the seismic records to be analyzed. These seismic records can span the entire year or, by using the "Time interval" block, specify the start and end times for analysis, expressed in Julian days. Selecting a specific filter type activates the five corresponding input value boxes, one for each value. The default value for the filter order is four. The default values for the envelope are: a) Window size = 50. b) Smoothing type = 3. c) Nan Tolerance = 1. An example of input these parameters is shown in Figure 7.

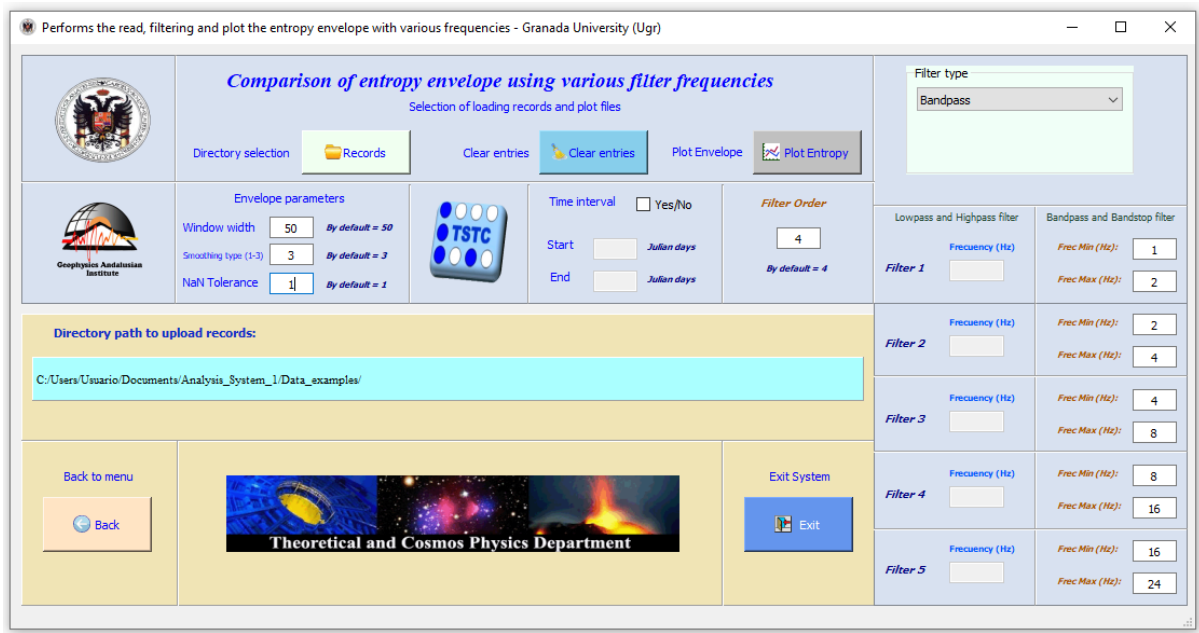

Figure 7: Results of the Module for reading, filtering, and comparing five values of a specific seismic record filter.

In the previous figure, it can be observed that the user has decided to set the default envelope parameters. They have chosen to use a band-pass filter to establish the five frequency (Hz) value intervals for the filters. These are: 1-2, 2-4, 4-8, 8-16, and 16-24.

Next, you only need to click the 'Plot Entropy' command button, and the graphical results will be displayed. If you do not agree with the selection, you can delete the entered data by clicking the 'Clear entries' button. A pop-up window will prompt you to confirm this choice. If you confirm, all entries will be deleted, and the default values will be restored. The selection of the data to be analyzed is delimited by the folder where the data is located. These can cover a whole year (365 days). However, you can choose to perform an analysis within a specific interval, such as a month. By checking the 'Time interval' box and designating the start and end value of the analysis in Julian days.

NOTE: The supplementary documentation, as well as the software comprising this module, can be accessed and downloaded from the link provided in "Code availability section".

## 3. Results

### 3.1. Resulting graph of module 1

The calculation analysis process of the one and three components, shown in section 3.1, results in the graph figure 8 displayed below.

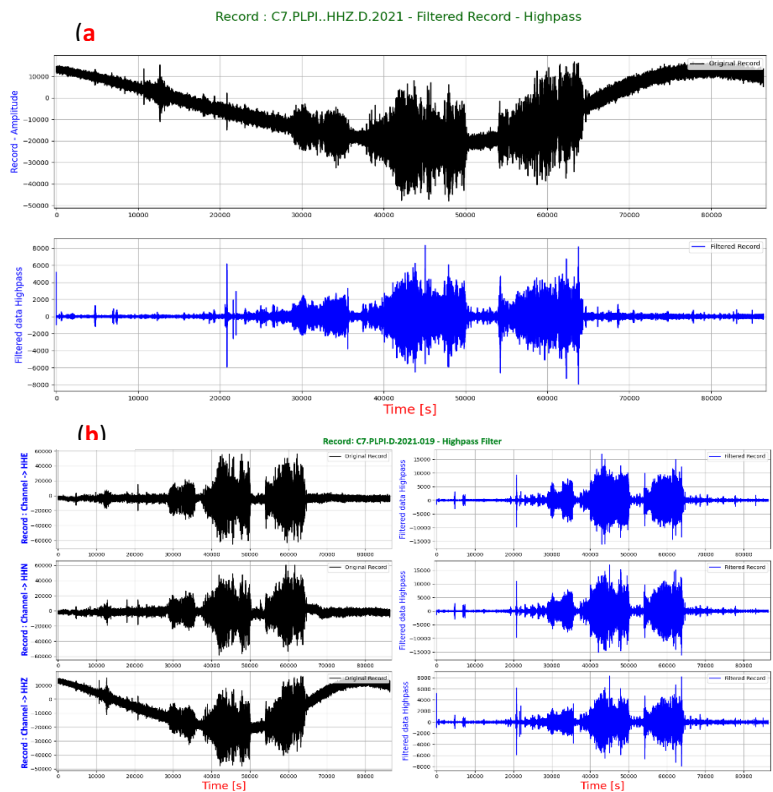

Figure 8. Graphical results of reading (a) One component and (b) of the three components of a seismic record of one day (24 hours), with input data shown in Figure 2. Using Module 1.

In the figure above, the graph (a) shows the result of the analysis with one component, and the graph (b) shows the result of the analysis with all three components, from top to bottom (East-West - EHW, North-South - NHW, and Vertical - HZ). The original signal is shown on the left, and the filtered signal is shown on the right. As mentioned, the graph can be stored in various formats, according to the user's preferences, for later use.

**4.3.** Resulting graph of module 2

The result of the calculation process of module two, presented in section 3.2, shows here a year of seismic analysis where the four main types of graphs generated during the three periods of state of the Cumbre Vieja volcano (pre-eruptive, eruptive and post-eruptive) of La Palma in Spain (January to December in the year 2021) are observed which correspond to the analysis of the parameters: 'Shannon entropy with its envelope', 'Entropy envelope only', 'Frequency index with its envelope', 'Kurtosis with the envelope' and 'Energy values with the envelope', respectively. This is shown in Figure 9, presented below.

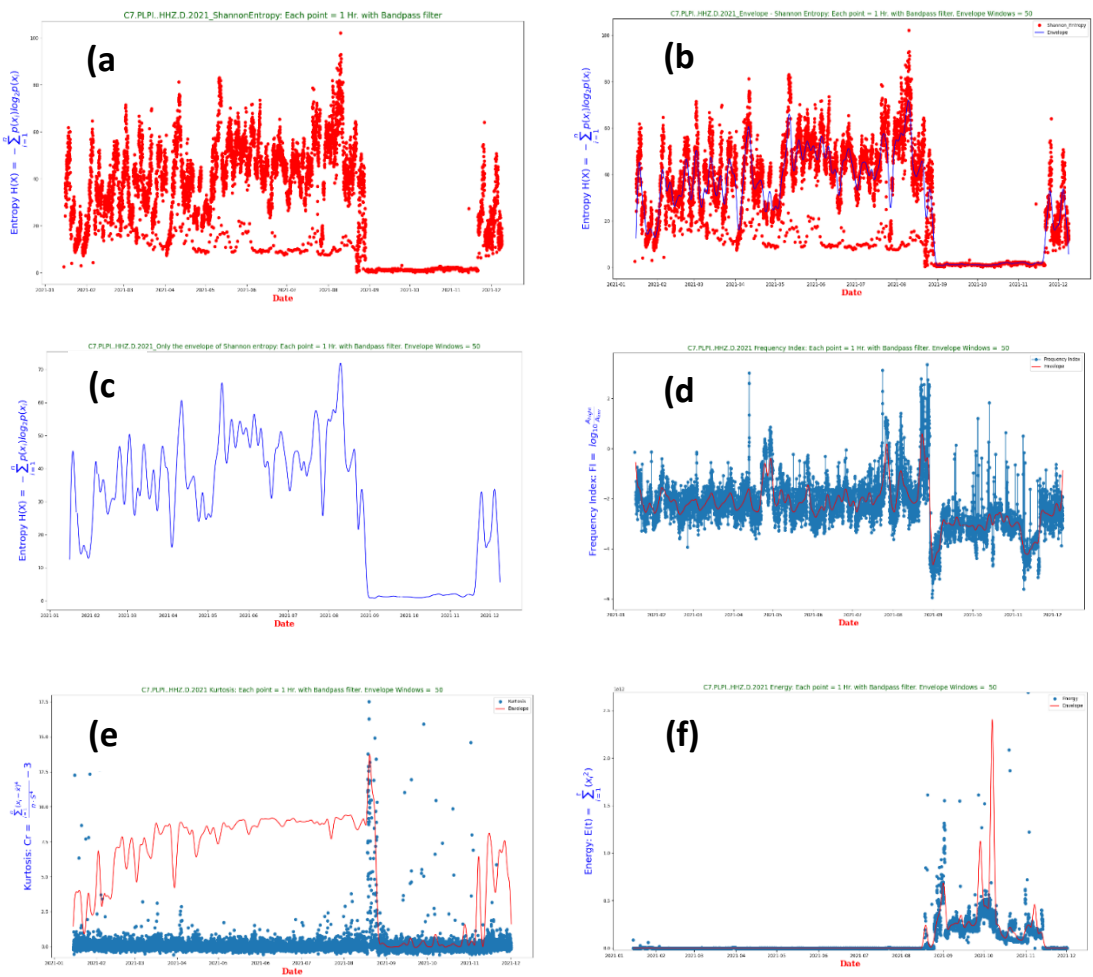

Figure 9. Graphical results obtained using Module 2 from one year of seismic records from the PLPI seismic station, using the vertical component (HHZ), filtering the signals with a BandPass filter (1 to 16 Hz). The user-assigned parameters were: an analysis window length of 1 hour, and a window envelope value of 50. The resulting graphs are: (a) Shannon entropy, (b) Shannon entropy with its envelope at 50, (c) Representation of the Shannon entropy envelope curve, (d) Frequency index with envelope, (e) Kurtosis values of seismic records with their envelope, and (f) Energy values with envelope.

The figure above shows the eruption that occurred on the volcano from September to December 2021. As the Shannon entropy values decay almost to zero, the frequency index and kurtosis, as expected during the eruptive phase, show high values, and the energy exhibits high-frequency peaks. All resulting values are represented by points according to the user's selection. Thus, each point on the graph represents 5 or 10 minutes, one hour, or 24 hours, depending on the selection (in this case, one hour). The envelope is represented by smoothed curves on a line, also according to the user's choice. By default, the value is set to 50 (but higher values, such as 100, 200, 300, etc., can be specified). This would smooth the dotted line but generate larger extreme values. The user must decide, based on the results obtained, which of these values best defines the signal's envelope. The values before and after the eruption are observed. It is also possible to define the time interval in which the drop in Entropy values becomes definitive, allowing for the timely establishment of the alert level. In a real-time system, this would be extremely important for determining both the start and end of the eruption, when the eruptive system in the volcano reaches a state of rest.

On the other hand, the module also offers the possibility of modifying the Shannon Entropy envelope (along with the other three metrics) to smooth the analysis line or curve of said envelope. The user can determine the parameters using various window sizes, for example, from 50 to 3000. However, the user must evaluate and decide which size to use based on the expected results. A larger window size significantly smooths the slope of the envelope, but also means that various peaks or jumps in the envelope may appear in parts of the signal, depending on the number of records present in a given year. This can be observed in Figure 10.

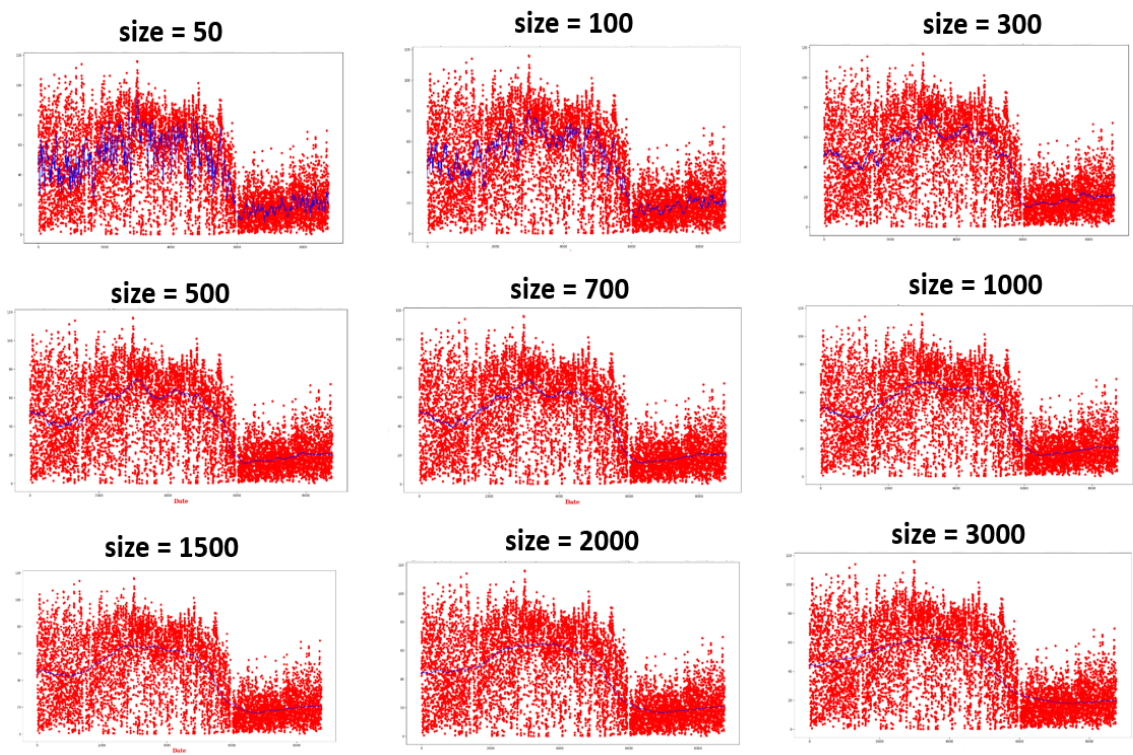

Figure 10. Results of a comparative graph of Shannon Entropy and its envelope, over a year of seismic records, showing various envelope window sizes such as: (*50, 100, 300, 500, 700, 1000, 1500, 2000 y 3000*).

The result of a comparative analysis using various window sizes: 10 minutes, one hour, and 24 hours, as well as the use of the four available filter types (*Low-Pass, High-Pass, Band-Pass, and Band-Stop*), can be observed in Figure 11 in the following image.

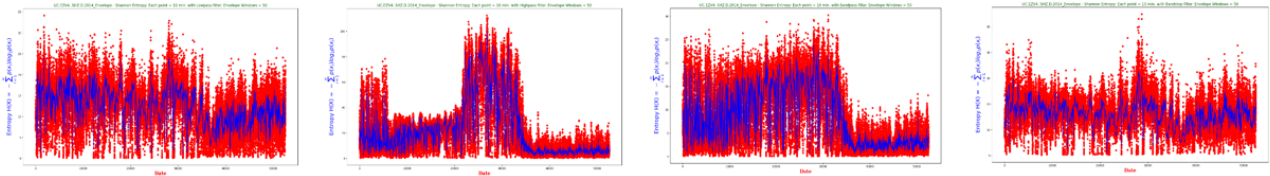
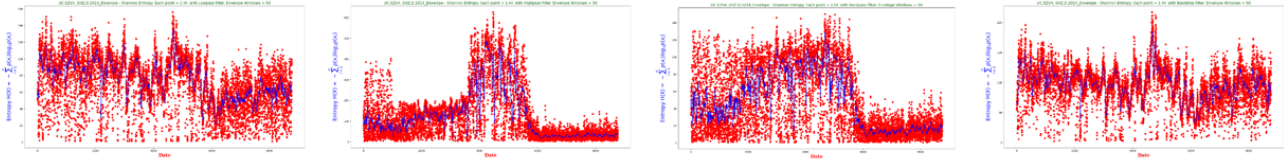
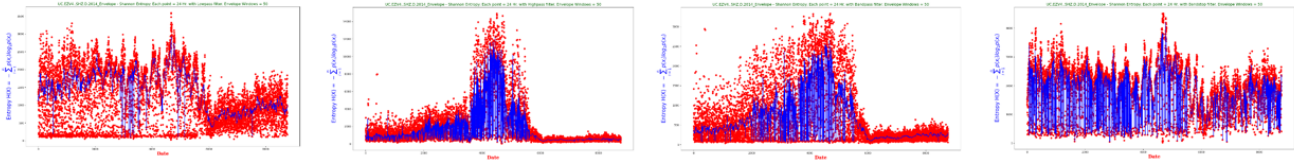

1) Lowpass Filter = 2 Hz.  2) Highpass Filter = 8 Hz.  3) Bandpass Filter = 1-15 Hz.  4) Bandstop Filter = 1-15 Hz.

Figure 11. Resulting graph of Shannon Entropy and its envelope, over a year of seismic records, from the comparison of using various analysis windows (*10 minutes, one hour, 24 hours*) with the four types of filters (*Lowpass, Highpass, Bandpass, and Bandstop*).

For example, the graphical result of the calculation of the probabilistic metrics available in the module (Shannon Entropy, Kurtosis Frequency Index and Energy), during a period of two consecutive years of seismic records (2021 to 2022), where the different scenarios of the evolution in the state of the volcano (pre-eruptive, eruptive and post-eruptive) can be observed more clearly, is presented in Figure 12, where the eruptive period has been demarcated between two green lines.

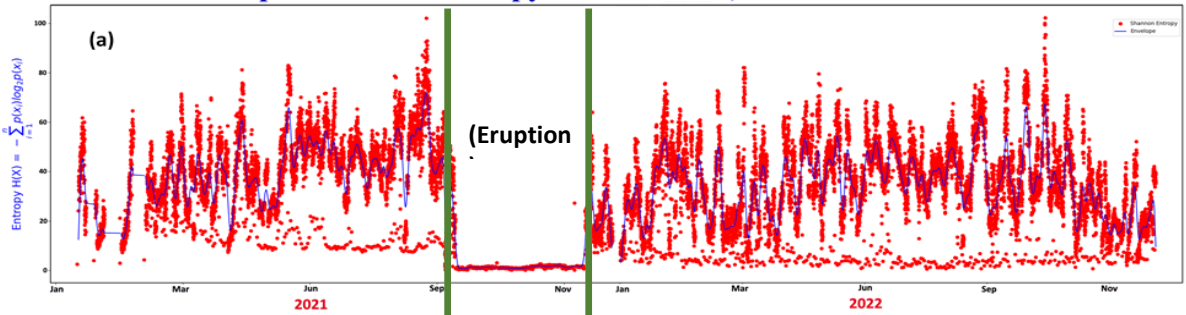
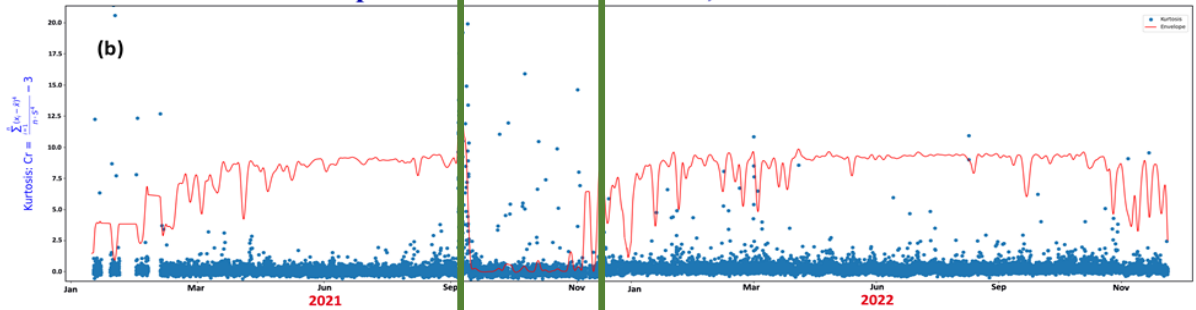
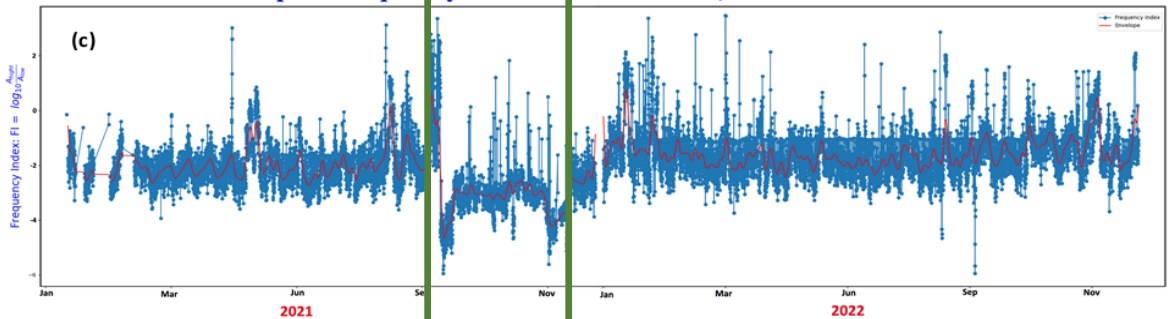
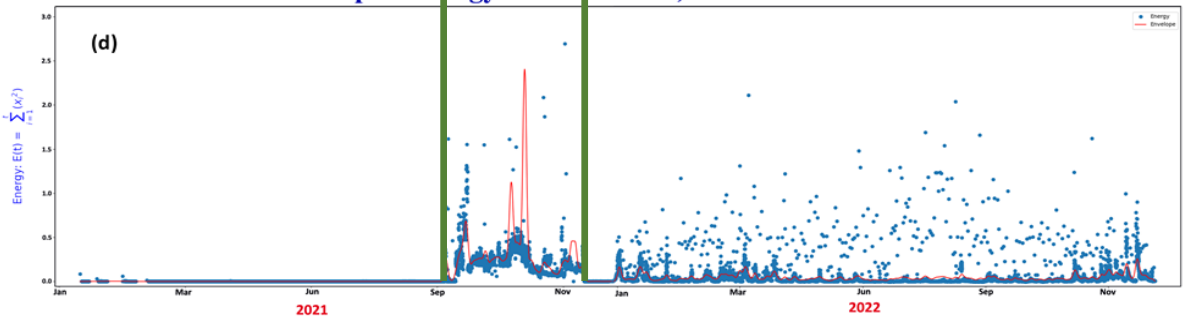

Figure 12. Graphical results using Module 2, from the analysis of the four metrics (Shannon Entropy, Kurtosis, Frequency Index, and Energy) with their respective envelopes of seismic records from the PLPI station of the La Palma volcano, during the activity observed in the period 2021 and 2022. The eruption period from September to December 2021 is shown, as well as the pre-eruptive and post-eruptive periods. The green line indicates the interval of the eruptive period.

Therefore, the expert criterion of the user, should prevail in deciding among the input parameters: which analysis window size, which type of filter, and which envelope window size to select, in order to achieve the best and most efficient analysis processes and thereby establish conclusions leading to an optimal final result.

**4.3.** Resulting graph of module 3

The calculation process of module three, presented in section 3.3, displays a graph corresponding to the analysis of parameters from five different filter intervals according to the type selected by the user. This is shown in Figure eleven below.

The graphic result of analyzing two consecutive years of seismic records, where the drop in Shannon Entropy is observed during an eruptive period, is seen in figure 13 below.

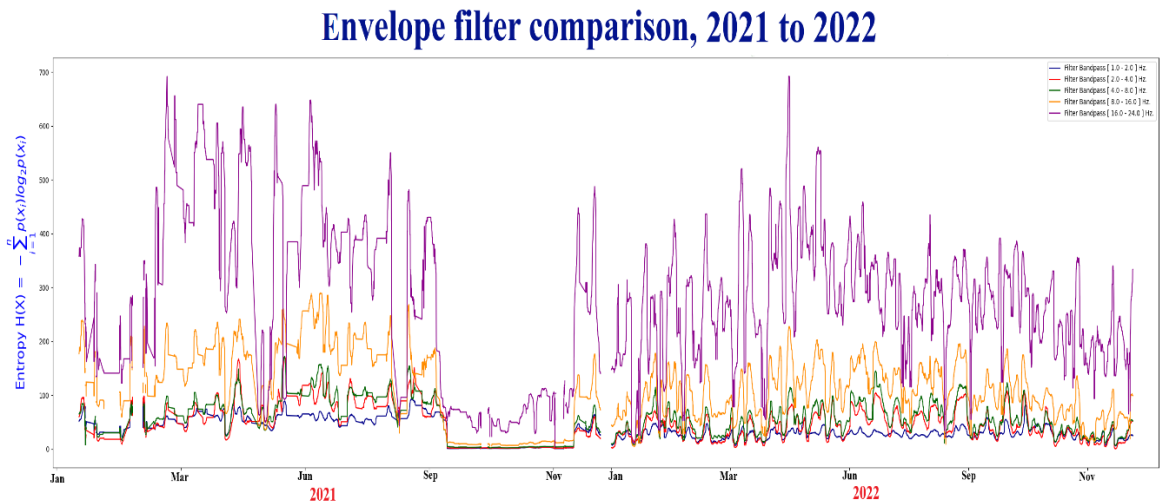

Figure 13. Graphical results of seismic records using Module 3. It presents the comparison of five band-pass filter interval values for seismic records spanning two years (2021 to 2022).

With the results observed in the previous graph, the user must decide which type of filter and which values to use to obtain the most efficient results for their analysis. To decide whether high or low frequencies present in the envelope better describe the jumps in entropy to determine explosions and earthquakes before or after a volcanic eruption, in order to quantify early warning states with greater accuracy.

## 4. Conclusions

The probabilistic metrics analysis and calculation system we have proposed and developed in this work for characterizing seismic-volcanic signals in active volcanoes is designed to be a set of efficient and reliable tools that are easy to use, access, understand, and implement in real-time monitoring systems, regardless of computational resources. It comprises intuitive interfaces that are simple to learn and manage, offering reliable technological assistance to the human operator in the process of reading and representing seismic records from active volcanoes, using one or three components (North-South, East-West, and Vertical), with the aid of digital filters. It facilitates the calculation and analysis of four important probabilistic metrics for analyzing the eruptive and predictive state of a volcano: Shannon Entropy, Kurtosis, Frequency Index, Energy, and their envelopes. It also allows for the comparison of Shannon Entropy envelopes with different filter values to determine the most suitable one for seismic signal analysis. The seismic analysis tool proposed here can be used and implemented in real time in studies and training for technicians at observatories and scientific research institutes. This would allow them to contribute added value to the creation of better databases, leading to improvements in the development of more efficient models for the recognition and training of early warning systems for future volcanic eruptions worldwide. Furthermore, it can also be included as educational material for the professional development of personnel responsible for seismic signal analysis at observatories and seismic institutes.

5. Acknowledgments

This software has been funded by:

a) This work has been funded by the Spanish Project PID2022-143083NB-I00, "LEARNING", funded by MICIU/AEI/10.13039/501100011033 and by FEDER, UE.

b) JMI and LG were partially funded by the Spanish project PROOF-FOREVER (EUR2022.134044)

c) PRD was funded by the Ministerio de Ciencia e Innovación del Gobierno de España (MCIN), Agencia Estatal de Investigación (AEI), Fondo Social Europeo (FSE), and Programa Estatal de Promoción del Talento y su Empleabilidad en I+D+I Ayudas para contratos predoctorales para la formación de doctores 2020 (PRE2020-092719).

d) PLEC2022-009271""DigiVolCa"", funded by MCIN/AEI, funded by MCIN/AEI/10.13039/501100011033 and by EU «NextGenerationEU/PRTR», 10.13039/501100011033.

**Code availability section**

Name of the code/library: [Shannon Entropy Estimator for the Characterization of Seismic-Volcanic Signals using Python](#)

Contact: e-mail and phone number: ligdamis@ugr.es /ligdamis@yahoo.com (+34)63951208

Hardware requirements: Desktop or laptop PC with Windows, Linux or Mac System

Program language: Python

Software required: Python Ver. 3.6 until 3.10

Program size: 18.1 MB (code and documentation). Data Base Colima 27.4 GB. Data Base La Palma 7.3 GB

Publicly available datasets were analyzed in this study. It is important to highlight that the raw data corresponding to the Colima and La Palma volcanoes analyzed for this study can be found online at ZENODO: Colima [https://zenodo.org/records/10781903], La Palma [https://zenodo.org/records/10781515].

The open source code developed in this work can be downloaded at **Github**

(https://github.com/LigA2024/Shannon-Entropy-Estimator-for-the-Characterization-of-Seismic-Volcanic-Signals-using-Python).

**References.**

**List of Figures**

Figure 12. Graphical results using Module 2, from the analysis of the four metrics (Shannon Entropy, Kurtosis, Frequency Index, and Energy) with their respective envelopes of seismic records from the PLPI station of the La Palma volcano, during the activity observed in the period 2021 and 2022. The eruption period from September to December 2021 is shown, as well as the pre-eruptive and post-eruptive periods. The green line indicates the interval of the eruptive period.

Figure 13. Graphical results of seismic records using Module 3. It presents the comparison of five band-pass filter interval values for seismic records spanning two years (2021 to 2022).